\documentclass[aps,prb,twocolumn,showpacs,superscriptaddress,groupedaddress]{revtex4}  
\usepackage{graphicx}  
\usepackage{dcolumn}   
\usepackage{bm}        
\usepackage{amssymb}   

\usepackage{amsmath}    
\usepackage{color}      
\usepackage{subfigure}  

\usepackage{ulem}  

\newcommand{\Fz}{F_\mathrm{Z}} 
\newcommand{\lSAW}{\lambda_\mathrm{SAW}} 
\newcommand{\fSAW}{f_\mathrm{SAW}} 
\newcommand{\vSAW}{v_\mathrm{SAW}} 
\newcommand{\Prf}{P_\mathrm{RF}} 
\newcommand{\Vbias}{V_\mathrm{BIAS}} 

\begin{document}

\widetext


\title{Scalable interconnections for remote exciton systems}


\author{S. Lazi\'c}
\email[e-mail address: ]{lazic.snezana@uam.es}
\affiliation{Paul-Drude-Institut f\"ur Festk\"orkperelektronik, Hausvogteiplatz 5-7, 10117 Berlin, Germany}
\affiliation{Departamento de F\'isica de Materiales, Universidad Aut\'onoma de Madrid, 28049 Madrid, Spain}
\author{A. Violante}
\affiliation{Paul-Drude-Institut f\"ur Festk\"orkperelektronik, Hausvogteiplatz 5-7, 10117 Berlin, Germany}
\author{K. Cohen}
\affiliation{Racah Institute of Physics, The Hebrew University of Jerusalem,  Jerusalem 91904, Israel}
\author{R. Hey}
\affiliation{Paul-Drude-Institut f\"ur Festk\"orkperelektronik, Hausvogteiplatz 5-7, 10117 Berlin, Germany}
\author{R. Rapaport}
\affiliation{Racah Institute of Physics, The Hebrew University of Jerusalem,  Jerusalem 91904, Israel}
\author{P. V. Santos}
\affiliation{Paul-Drude-Institut f\"ur Festk\"orkperelektronik, Hausvogteiplatz 5-7, 10117 Berlin, Germany}

\date{\today}

\begin{abstract}
Excitons, quasi-particles consisting of electron-hole pairs bound by the Coulomb interaction, are a potential medium for processing of photonic information in the solid-state. Information processing via excitons requires efficient techniques for the transport and manipulation of these uncharged particles. We introduce here a novel concept for the interconnection of multiple remote exciton systems based on the long-range transport of dipolar excitons by a network of configurable interconnects driven by acoustic wave beams. By combining this network with electrostatic gates, we demonstrate an integrated exciton multiplexer capable of interconnecting, gating and routing exciton systems separated by millimeter distances. The multiplexer provides a scalable platform for the manipulation of exciton fluids with potential applications in information processing.
\end{abstract}

\maketitle  


\section{Introduction}
\label{Introduction}
The strong interaction with photons as well as with electronic excitations makes excitons ideal particles for the processing of photonic information. Within this approach, the information carried by photons is first converted to excitons, which are then manipulated by electric, optical, acoustic or magnetic fields and subsequently reconverted to photons for further information transmission \cite{Ref1,Ref2}. Exciton manipulation profits from strong nonlinearity in excitonic systems mediated by their electric-dipole interactions, which can be explored for the realization of all-optical control devices \cite{Ref3}. In addition, the interconversion between photons and excitons can be carried out coherently, i.e., by transferring the phase information between the two systems. The latter becomes especially interesting when combined with macroscopically coherent phases of either exciton polaritons \cite{Ref4,Ref5} or indirect excitons (IX) \cite{Ref6,Ref7,Ref8}, which provide a new platform for exciton-based information processing. Yet, the coherent manipulation of these excitonic phases as well as their integration into optoelectronic circuits  remains a great challenge.

\begin{figure}
\includegraphics[width=\columnwidth, keepaspectratio=true]{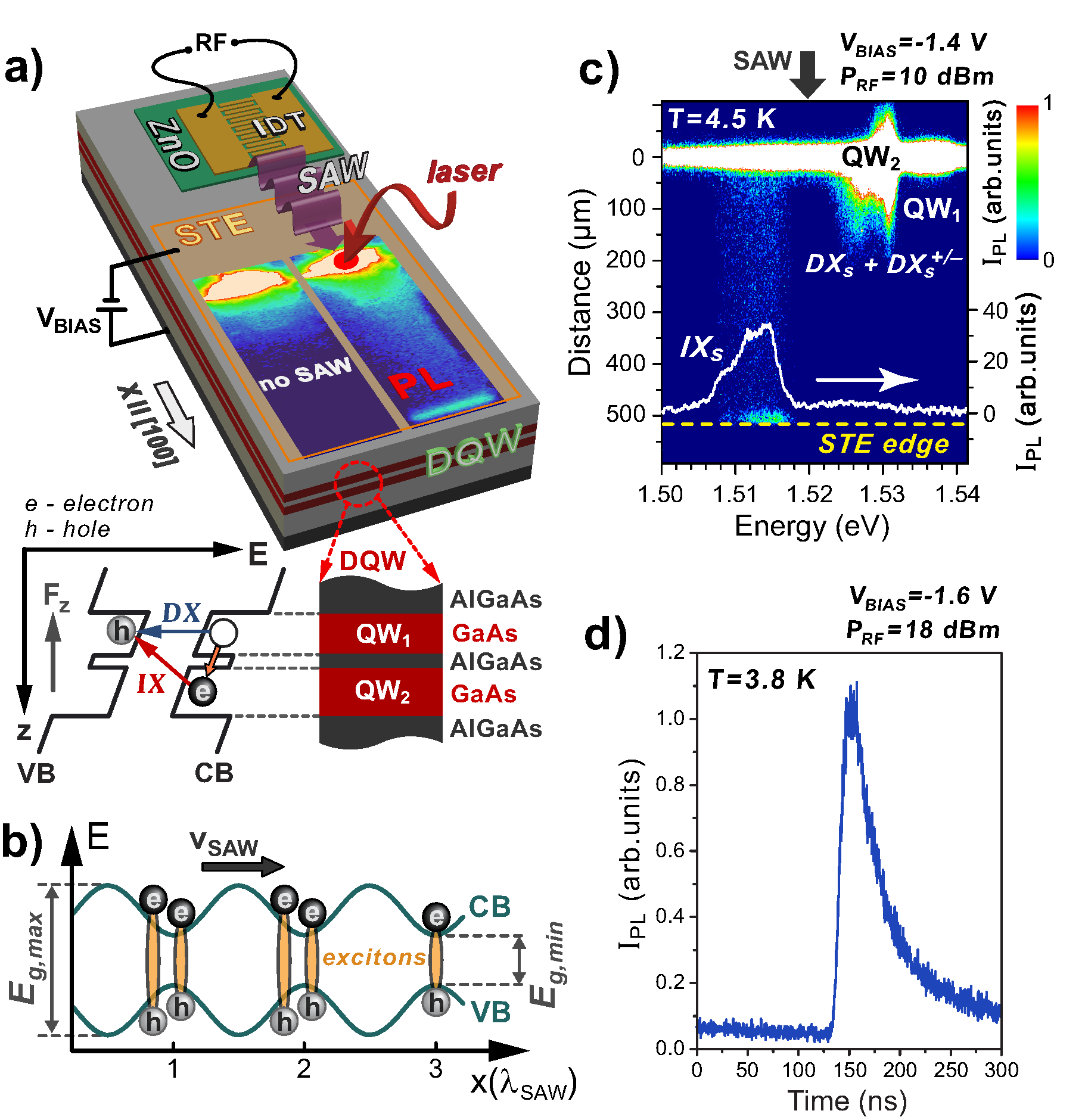}
\caption{(a) Sample layout: Indirect excitons (IXs) photoexcited in the GaAs/(Al,Ga)As DQW are transported by a surface acoustic wave (SAW) launched by applying an electric power P$_{RF}$ to an interdigital transducer (IDT) placed on a piezoelectric ZnO island. The SAW amplitude is stated in terms of the nominal RF power P$_{RF}$ applied to the IDT. The IX lifetime and energy are controlled by the bias voltage $\Vbias$ applied between the semitransparent top electrode (STE) and the doped substrate. The superimposed PL images compare the emission in the absence and presence of a SAW. Inset: Energy band diagram of a DQW along the growth (z) direction showing the direct (DX) and indirect (IX) exciton transitions.~(b) IX transport by the moving strain modulation of the conduction (CB) and valence band (VB) in a DQW.~E$_{g,min}$~(E$_{g,max}$) denotes minimum (maximum) bandgaps induced by SAW strain field along the x=[100] direction. (c) Spectral PL image of the transport channel at T=4.5~K.~The superimposed spectrum was recorded at the STE edge.~(d) Time-resolved PL trace recorded at T=3.8~K at the edge of the STE.}
\label{Fig1}
\end{figure}

Excitons are metastable with lifetimes normally in the \textit{ns}-range for conventional direct excitons (DXs) and in the \textit{ps}-range for exciton polaritons. Applications require particles with long lifetimes, which can be efficiently stored, manipulated and transported between different locations. Such a requirement is met by indirect excitons (IXs) in bilayer semiconductor systems consisting of two quantum wells (QWs) separated by a thin barrier, termed a double quantum well (DQW). As illustrated in the inset of Fig.~\ref{Fig1}(a), the application of a transverse electric field $\Fz$ forces the oppositely charged exciton constituents (i.e., electron and hole) into different QWs.~The spatial charge separation enhances the IX radiative lifetime by orders of magnitude (up to a $\mu$s-range) as compared to DXs, while still maintaining the Coulomb correlation between the electrons and holes.

Different studies have addressed the manipulation of IXs on time scales shorter than their lifetime. The trapping of IXs even down to the single particle level \cite{Schinner_PRL110_127403_13} was demonstrated using a variety of potential landscapes created by optical \cite{Ref9}, magnetic \cite{Ref10}, electric \cite{Ref11,Ref12} or strain \cite{Ref13} fields.~Applied electric fields, in particular, can control the IX energy and lifetime via the quantum confined Stark effect (QCSE) induced by $\Fz$ \cite{Ref14}.~Proposals for all-optical control via remote dipolar interactions of IX fluids have also been put forward \cite{Ref3}. Recently, reference \cite{Ref15} has introduced an exciton optoelectronic transistor, where the exciton flow (over a few $\mu$m) induced by potential energy gradients 
has been exploited for the realization of simple operations on photonic signals, such as directional switching and merging \cite{Ref16,Ref17}.~Several groups have also demonstrated the transport of IXs over longer distances by diffusion \cite{Ref18}, drift \cite{Ref19} as well as by moving electrostatic lattices \cite{Ref20}. Finally, we have recently  shown that IX can be efficiently transported over several hundreds of $\mu$m by the moving (and tunable) strain modulation induced by a surface acoustic wave (SAW) propagating on the DQW structure.\cite{Ref21} 



Previous reports have addressed the coupling of a reduced number of, in most cases, closely spaced IX systems. An interesting question is whether an approach can be envisaged to interconnect multiple remote IX sites, providing, therefore, a multi-functional and scalable technology for exciton-based  optoelectronic circuits. 
In this paper, we introduce a novel concept for the scalable interconnection of remote IX systems based on a network of acoustic beams and electrostatic gates. The feasibility of the concept is demonstrated by an exciton acoustic multiplexer (EXAM) - an integrated multiport device for the storage and  controlled transfer of IXs between any two nodes of a two-dimensional (2D) array of IXs systems. 
The EXAM consists of a network of I/O ports  interconnected by an array of transport channels driven by SAWs. 
Each I/O port can be isolated from the array of transport channels by an exciton acoustic transistor (EXAT). As a results, these ports can be used for storage of IXs as well as for their interconversion to photons. 
Scalability, presently a main limitation for quantum computational systems, is ensured by the sublinear dependence of the EXAM dimensions on the number of interconnected nodes. 
The experimental transport dynamics between the nodes is well-reproduced by a theoretical model of self-interacting IX fluids in a dynamic potential landscape, which successfully demonstrates the ability to design, simulate and test  complex IX devices.
The EXAM concept is extensive and compatible with the well-established acoustic transport and manipulation of quantum excitations by SAWs \cite{PVS152,PVS171,PVS265,NPSne,PVS223}.

In the following section (Sec. II), we describe the sample  fabrication  details as well as the optical techniques employed for the detection of IX transport and multiplexing. The experimental results are then presented in Sec.~\ref{Results}. 
Here, we start with investigations of the dynamics of IXs under acoustic fields  (Sec.~\ref{Acoustic_exciton_transport}), 
which demonstrate that SAW fields can capture IX packets and transport them  over large distances. 
We then address the control of IX fluxes, which is carried out using EXATs  (Sec.~\ref{SecEXAT}). 
In Sec.~\ref{SecEXAM}, the flow control by EXATs is then combined with long-range acoustic transport to realize an IX multiplexer. Finally, Sec.~\ref{Discussions_and_Conclusions}  discusses issues regarding the application of EXAMs as a building block for integrated electro-optical devices based on IXs and summarizes the main conclusions of this work.

\section{Experimental Details}
\label{Experimental_Details}
The experiments were carried out on an (Al,Ga)As field-effect heterostructure containing three sets of DQWs grown on a $n^+$-doped GaAs (001) wafer by molecular beam epitaxy. Each DQW consists of two GaAs QWs with thickness of 14 and 17 nm (QW$_{1}$ and QW$_{2}$, respectively - cf. lower right inset in Fig. 1 (a)) separated by a 4-nm thick Al$_{0.33}$Ga$_{0.67}$As tunnelling barrier. The DQWs were positioned at a depth of 700 nm below the sample surface within an undoped 2.5-$\mu$m-thick GaAs/(Al,Ga)As superlattice (SL).

The electric field $\Fz$ is induced by a reverse bias $\Vbias$ applied across the Schottky diode formed between a 10-nm-thick semitransparent titanium electrode (STE) on the sample surface and the n$^+$-doped substrate, which serves as the common ground [cf.~Fig.~\ref{Fig1}(a)]. The lower left inset shows the bending of the conduction (CB) and valence (VB) band edges along the growth (z) direction induced by $\Fz$. The excitons are locally excited underneath the STE area by a 780~nm pulsed diode laser (average power density of approx. 12~J/cm$^2$). The use of SLs reduces the leakage photo-currents to values below 1 nA (for bias between -2 and 2 V and laser excitation power in the range up to a few tens of $\mu$W), thus ensuring that IXs are only formed by photoexcited carriers. 

The Rayleigh SAWs (with wavelength $\lSAW$=2.8~$\mu$m corresponding to a SAW frequency $\fSAW =1$~GHz at helium temperatures) were electrically excited along a $<$100$>$ surface direction by applying a radio-frequency (RF) signal to interdigital transducers (IDTs) deposited on piezoelectric ZnO islands. These SAWs do not carry a piezoelectric field outside the IDT region, thus preventing field-induced exciton dissociation \cite{Ref21}.~Their strain field creates, via the deformation potential interaction, the moving type-I modulation of the CB and VB band edges of the DQW~\cite{Ref22} displayed in Fig.~\ref{Fig1}(b). Here, E$_{g,min}$ (E$_{g,max}$) denotes the minimum (maximum) bandgap at the SAW phases of largest tensile (compressive) strain. Photoexcited IXs are captured close to the regions of E$_{g,min}$, which move with the acoustic velocity $\vSAW$~\cite{Ref21}. The confinement of the electron and the hole at the same lateral (i.e., along the surface) coordinate prevents exciton dissociation, thus making this approach qualitatively different from charge transport by the type-II bandgap modulation induced by the piezoelectric SAWs. The depth of the DQW structures was selected in order to obtain a type-I potential with approximately equal modulation amplitudes for the conduction and valence bands.\cite{Ref23} 

The transport of IXs by the SAW strain field is tracked by imaging the photoluminescence (PL) emitted via exciton recombination along the SAW transport path. The laser excitation was focused onto a 5 $\mu$m (10 $\mu$m) spot on the top semitransparent electrode by a 10x (5x) microscope objective. The light emitted along the SAW transport channel was collected by the same objective, dispersed by a single grating spectrometer and imaged onto a cooled charge coupled device (CCD) camera with spatial resolution of 1~$\mu$m (2~$\mu$m). 

\section{Results}
\label{Results}

\subsection{Acoustic exciton transport}
\label{Acoustic_exciton_transport}

The two PL images superimposed on the device structure of Fig.~\ref{Fig1}(a) compare the excitonic PL  in the absence (left plot) and presence (right plot) of a SAW. While in the former the emission is restricted to the region near the excitation spot, the latter shows substantial PL at the edge of the STE located approx. 500~$\mu$m away from the laser spot. The remote PL is attributed to the recombination of IXs transported by the SAW at the edges of the STE, where the sudden reduction of the vertical field $\Fz$ creates a potential barrier blocking the IX transport \cite{Ref12}.~As the IXs are pushed by the SAW against this barrier their concentration increases while the lifetime reduces, thus resulting in a stronger PL.

The long-range transport of IXs is confirmed by the spectral dependence of the PL along the transport channel displayed in Fig.~\ref{Fig1}(c). The PL  around 1.53~eV is dominated by the recombination of spatially direct neutral (DXs) and charged (DX$^{\pm}$) exciton complexes associated with the two QWs. This emission extends to distances of up to approximately 150~$\mu$m from the generation point due to the expansion of hot excitons and free carriers created by the pulsed laser \cite{Ref24}. In contrast, the remote emission near the STE edge only shows the red-shifted PL signal, which is attributed to the recombination of acoustically transported IX packets. The energy of this remote PL is determined (via the QCSE) by the bias applied to the STE. The PL intensities integrated over the transport channel with and without a SAW differ by less than 1\%, demonstrating that the SAW transports the excitons introducing negligible losses. Moreover, by calculating the ratio between the integrated PL intensity for IXs at the excitation point in the absence of SAW and at the recombination point for P$_{RF}$=10 dBm (cf. Fig.~\ref{Fig1}(c)), one can estimate the exciton transport efficiency, which for a distance of 500 $\mu$m is around 50\%.


The dynamics of the IX transport was probed by time-resolved PL measurements at the end of the SAW transport channel, as illustrated in Fig.~\ref{Fig1}(d). From the delay of the PL pulse onset we calculate a maximum propagation velocity of 2.4$\pm$0.1~$\mu$m/ns, which is close to the SAW group velocity of $\vSAW$=2.6$\pm$0.1~$\mu$m/ns. The width of the PL pulse is determined by the  initial size of the IX cloud created by the pulsed laser excitation. We observe a slight broadening of the PL profile as well as the development of a tail with increasing transport distances.  These features, which will be addressed in details elsewhere,\cite{Ref25} are attributed to trapping sites along the transport path induced by potential fluctuations, which capture IX and delay their propagation. 
The transport within well-defined $\mu$m-sized packets over long distances and at constant velocity contrasts with conventional diffusion \cite{Ref18} or drift-driven \cite{Ref15,Ref16,Ref17,Ref19,Ref20} processes, where the IX fluid undergoes self-expansion. In addition, it enables synchronization with pulsed optical excitation  
by appropriately selecting the SAW frequency.

\subsection{Acoustic exciton transistor - EXAT}
\label{SecEXAT}

\begin{figure}
\includegraphics[width=\columnwidth, keepaspectratio=true]{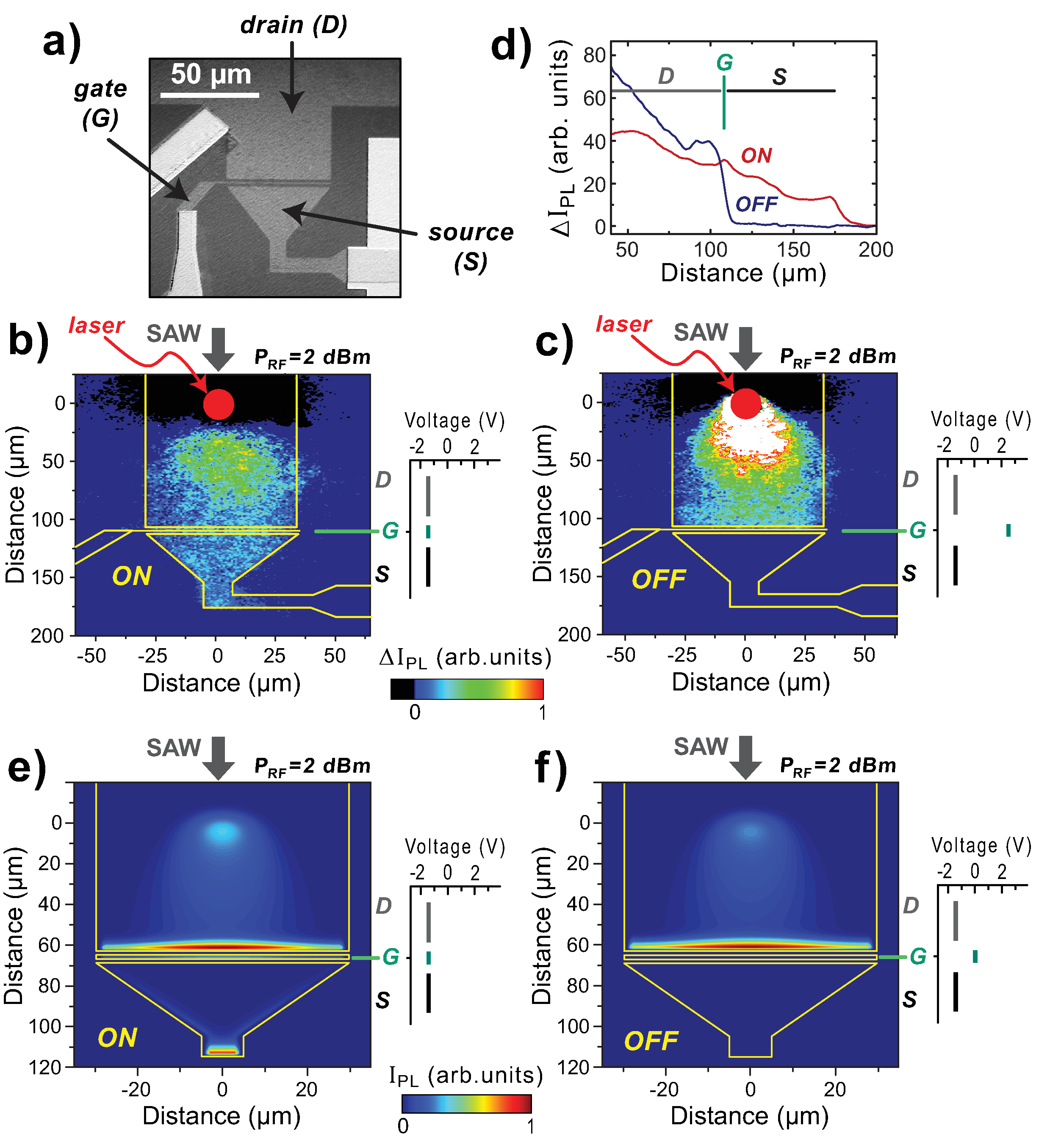}
\caption{(a) Top view of the EXAT showing the source (S), drain (D) and gate (G) electrodes. The G electrode is 1.5 $\mu$m wide and separated by the same distance from the adjacent S and D regions. The lower color plots display PL images of the acoustic transport of IXs across the EXAT recorded in the (b) ON state (with biases V$_{S}$=V$_{D}$=V$_{G}$=-1.4 V, cf. right inset) and (c) OFF state (V$_{S}$=V$_{D}$=-1.4 V, V$_{G}$=2.5 V, cf. left inset). In both cases, the IXs were driven by a SAW generated using P$_{RF}$=2 dBm. The filled circles mark the position of the laser excitation spot. (d) Spatially resolved $\Delta$I$_{PL}$ profiles along the transport channel in the ON (red curve) and OFF (blue curve) states. (e) and (f) display the corresponding time-averaged PL profiles calculated by a numerical model (see text). The overlays indicate the positions of drain, gate and source electrodes.}
\label{EXAT}
\end{figure}

The SAW-driven long-range IX flow can be efficiently controlled by the EXAT transistor placed on the transport path. The EXAT electrode configuration (cf. Fig.~\ref{EXAT}(a)) is similar to the exciton optoelectronic transistor introduced in Refs.\cite{Ref15,Ref16,Ref17}. Here, the voltage applied to the gate electrode (G) creates (via the QCSE) a tunable energy barrier, which controls the acoustically driven IX flow between the source (S) and the drain (D) electrodes. In contrast to its optoelectronic counterpart, the IX flow through the EXAT is acoustically (rather than diffusion or drift-) driven with the flow direction defined by the SAW propagation direction. Note that, in an EXAT, the roles of the source and drain electrodes are interchangeable simply by reverting the SAW propagation direction (not shown). To simplify the notation, we denote by $"source"$ the port with the smaller area. 
The switching action of the EXAT is demonstrated by the differential PL ($\Delta$I$_{PL}$) images in Figs.~\ref{EXAT}(b) and \ref{EXAT}(c). These images display the changes induced by the SAW and were obtained by subtracting PL images recorded in the presence and absence of the acoustic excitation.  In both cases, excitons were optically excited on D, which was kept at the same bias voltage as S (V$_{S}$=V$_{D}$=-1.4 V). Note that the negative $\Delta$I$_{PL}$ values around the excitation spot are due to the IX extraction by the SAW and their transport along the SAW propagation direction. In the ON state (Fig. 2(b)), IXs can be easily transferred from D to S by applying a gate voltage V$_{G}$=V$_{S}$=V$_{D}$ (cf. right inset of Fig.~\ref{EXAT}(b)). The fringe electric field at the edges of the electrodes induce a potential barrier for IXs between S and G and between G and D, even when all electrodes are at the same potential. This barrier, however, is low in the ON state and can be easily surmounted by the acoustic field. The transport, however, becomes completely blocked when the gate voltage is increased to V$_{G}$=2.5 V (OFF state in Fig. 2(c)). The switching efficiency is quantified in Fig. 2(d), which compares $\Delta$I$_{PL}$ profiles along the SAW transport channel in the OFF and ON states. Note that in contrast to purely electrostatic transistor \cite{Ref15,Ref16,Ref17,Ref18}, which require a potential gradient between source and drain, the EXAT can be operated with these two ports at the same bias levels. 
The lower panels in Fig.~\ref{EXAT} ((e) and (f)) compare the experimental results for the IX transport across an EXAT with a numerical modelling of the dynamics of the IX fluid. This model, which uses the experimental sample parameters, includes the mutual interactions between IXs and the external potential induced by the electrostatic gates and the SAW. Details of the model together with addition simulation results (including videos demonstrating the EXAT operation) are contained  in the supplementary material SM accompanying the manuscript. The numerical analysis of the IX transport across the EXAT described in the SM  show that the effect of the source, gate and drain edges on the exciton flow are negligible when the EXAT is in the ON state.  The good agreement between the experiment and modeling confirms that the device functions are well-predicted by our model of IX fluid dynamics in mobile potentials. 

The electrically controlled separation between the input and output ports (i.e., S and D electrodes) in the EXAT allows us to prove that the acoustic field really transports both exciton constituents (i.e., both electrons and holes) along the DQW structure. This question is relevant since the remote PL in IX transport experiments under a single STE can also be induced by the recombination of one type of carrier transported along the semiconductor channel with carriers of opposite polarity injected from the contacts (as for the IX bright spots reported in \cite{Ref8}). 
In order to address this issue, we have determined the absolute sensitivity of the setup, which is defined as the photon flux required for a given count rate measured on the CCD camera detecting the PL. The sensitivity calibration was carried out using a reference laser beam reflected on a metal film deposited on the sample surface. Knowing the reflectance of the metal, it is possible to estimate the absolute reflected flux of photons. The ratio between the photon flux collected by the microscope objective in front of the sample and the CCD count rate gives the required calibration coefficient. 

The number of photons generated via the recombination of IXs was determined from the measured CCD count rate by taking into account the collection angle (which depends on the refractive index of the sample layers and on the numerical aperture of the objective lens) and the transmission of the semi-transparent metal electrodes. If the PL signal comes from the recombination of one type of carriers transported by the SAW with carriers of opposite polarity electrically injected from the electric contacts, we expect the contact current to be given by $I_c=e N_\mathrm{ph}$, where $N_\mathrm{ph}$ is the detected photon flux and $e$ the electron charge. If the measured current is much less than $I_c$, then at most of the electrons or holes forming the IXs must have been transported along the DQW structure by the SAW. For the experimental conditions of Fig.~2b, where the laser beam is placed on the drain electrode, the measured photon flux in the source of the EXAT would require a source current $I_c \approx 50$~nA. This current is two orders of magnitude larger than the measured source current (of 0.5~nA) under the same illumination conditions, showing that both electrons and holes must have been transported by the SAW. Similar results were obtained for the current in the drain electrode, where recombination takes place at the edge of the drain area. 

%

\subsection{Acoustic exciton multiplexer - EXAM}
\label{SecEXAM}

\begin{figure}
\includegraphics[width=\columnwidth, keepaspectratio=true]{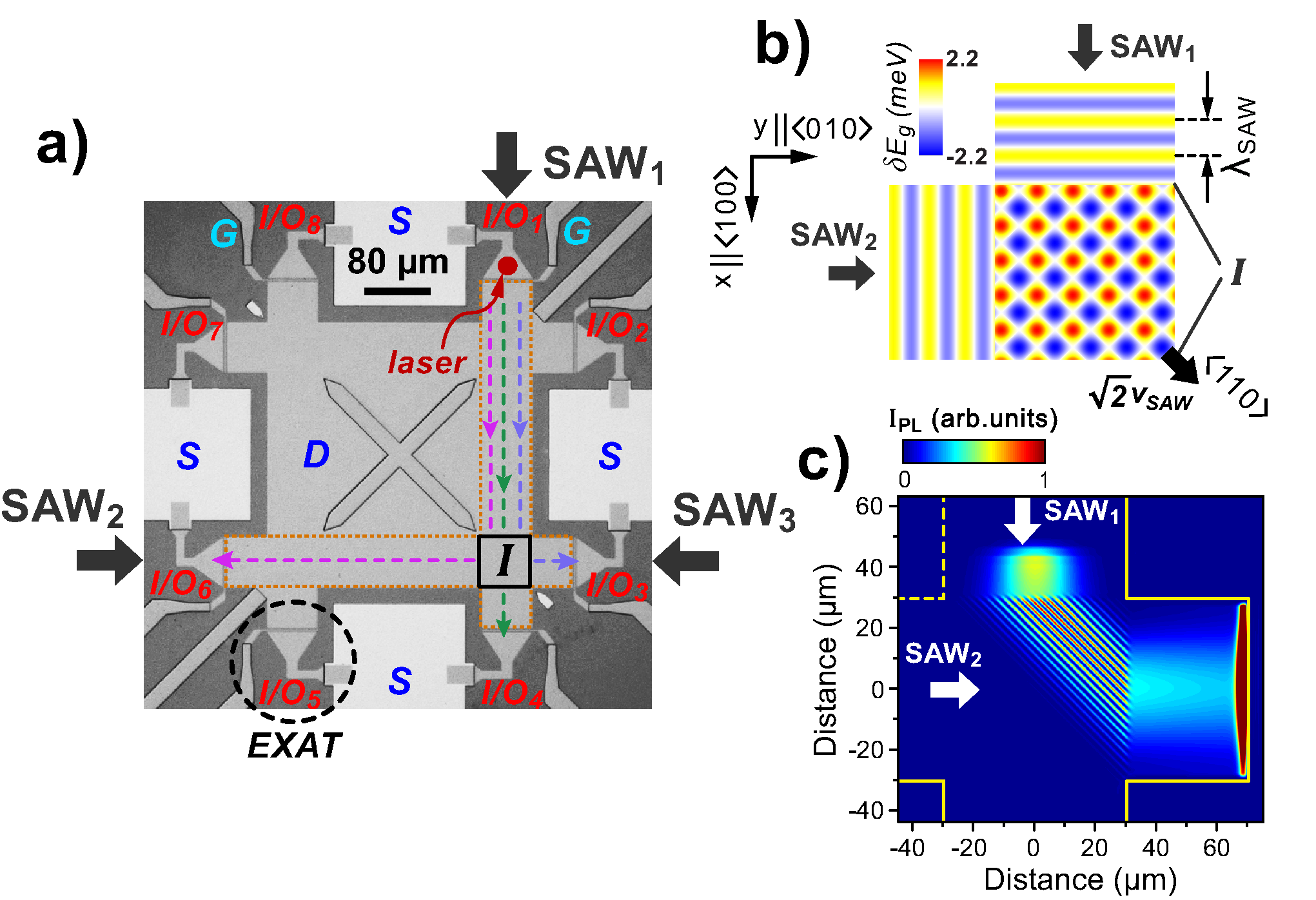}
\caption{(a) Top view of the acoustic exciton multiplexer (EXAM): each I/O area contains an exciton optoelectronic transistor (EXAT) and an IDT for SAW generation (not shown). The vertical (horizontal) dotted rectangle delineates the  transport channel formed by SAW$_1$ (SAW$_2$ or SAW$_3$) underneath the common drain. The dashed arrows indicate IX transfer between SAW$_1$, SAW$_2$ and SAW$_3$. The solid square labeled \textit{I} denotes the intersection region of SAW$_1$ and SAW$_2$ (or SAW$_1$ and SAW$_3$). (b) Array of moving potential dots formed by SAW interference at \textit{I}. (c) Simulated time-averaged emission pattern within \textit{I} showing the IX transfer between SAW$_1$ and SAW$_2$.}
\label{Fig2}
\end{figure}

We now introduce the EXAM (cf.~micrograph of Fig.~\ref{Fig2}(a)), a device capable of routing IX fluids between different input/output ports ($I/O_{i}$, labeled by the port index i=1,...,8).~Each I/O port contains an IDT (not shown) to generate an outgoing SAW beam as well as an exciton acoustic transistor (EXAT, cf. Fig.~\ref{EXAT}(a)) to control the IX flux between the EXAT source area and the common drain region, where multiplexing takes place. In each I/O port, an input optical signal (represented by a laser incident on the source electrode of EXAT$_1$ in Fig.~\ref{Fig2}(a)) can be converted into IXs and stored in the EXAT source electrode. Alternatively, stored IXs can be reconverted to photons by selecting an appropriate source voltage to reduce the IX recombination lifetime. The EXAT source acts, therefore, as a voltage-tunable location for photon-to-exciton interconversion as well as for IX storage. 

IX transfer between two facing I/O ports (e.g., ports 1 and 4 in Fig.~\ref{Fig2}(a)) is carried out by simply turning the corresponding SAW and EXAT (i.e., SAW$_1$ and EXAT$_1$) ON. The multiplexing function relies on the transfer of IXs between two orthogonal acoustic beams (e.g., from the vertical beam SAW$_1$ to the horizontal beam SAW$_2$ or SAW$_3$) at their intersection area (\textit{I} in Fig.~\ref{Fig2}(a)). We have previously shown that the moving piezoelectric dots can efficiently transport uncorrelated electron-hole pairs between orthogonal piezoelectric SAWs \cite{Ref26}. A similar concept using a moving array of strain dots is applied here to transfer IXs between two non-piezoelectric SAWs. The moving dot array is created by the interference of the two SAW beams. As seen in Fig.~\ref{Fig2}(b), the dots propagate with a velocity $\sqrt{2}\vSAW$ ($\vSAW$ denotes the SAW phase velocity) in channels forming an angle of 45$^\circ$ with respect to the propagation directions of the two interfering SAWs \cite{Ref26}. Due to their oblique propagation direction, the dots can capture the IXs from the input beam and transfer them to the other beam, as illustrated in the time averaged PL simulation of Fig.~\ref{Fig2}(c). The oblique lines show the propagation path of moving dots: a detailed numerical analysis of the transfer process including videos is presented in  the SM. 

\begin{figure}
\includegraphics[width=\columnwidth, keepaspectratio=true]{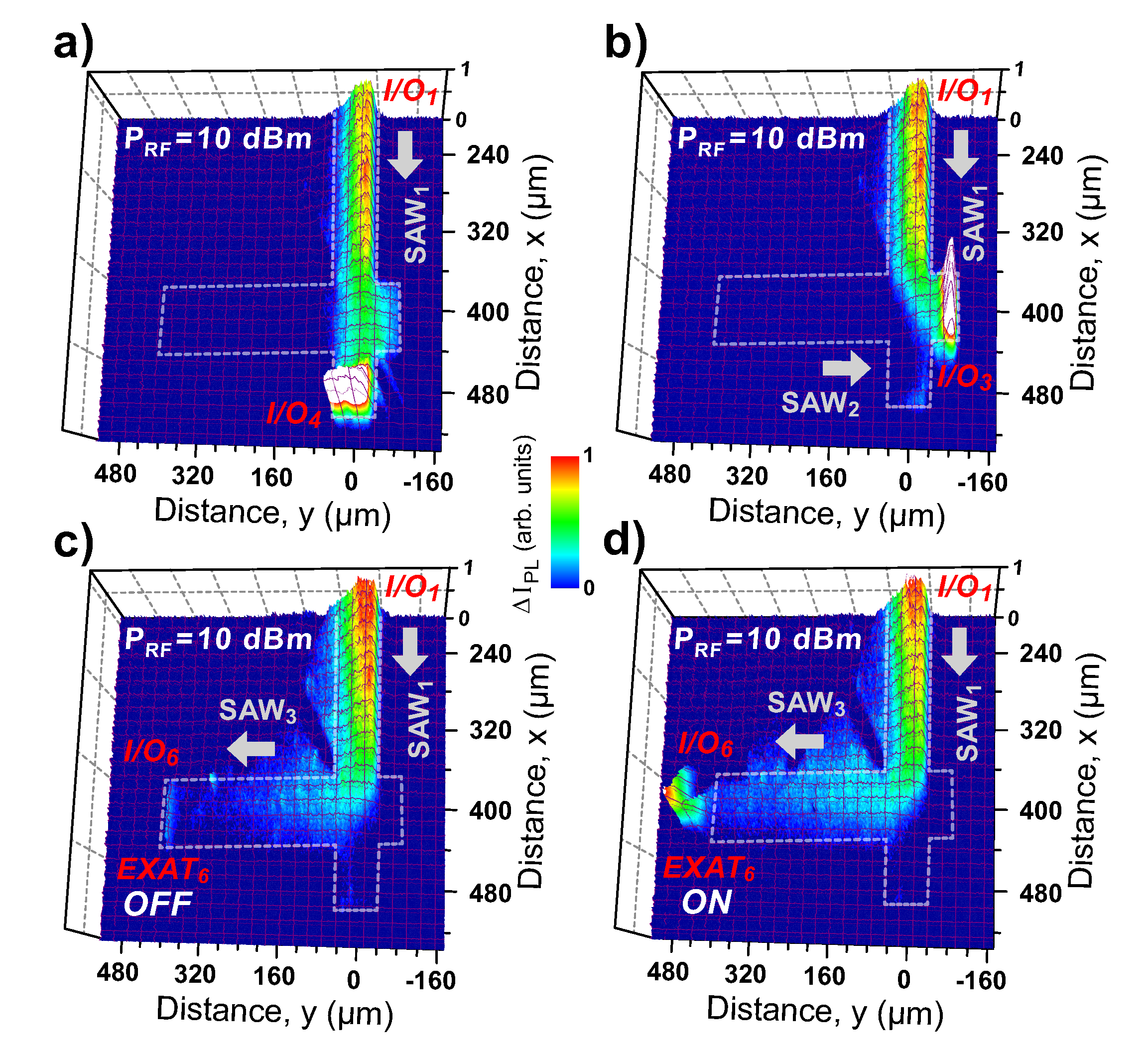}
\caption{PL images of the acoustic IX transfer (a) along SAW$_1$ and between (b) SAW$_1$ and SAW$_2$ or (c) SAW$_1$ and SAW$_3$. The output EXAT in (a)-(c) is in the OFF state.~(c),(d) Control of IX flux by the gate electrode of EXAT$_6$.~All SAW beams are generated by applying $\Prf$=10~dBm to the IDTs. The vertical (horizontal) dotted rectangle marks the transport channel formed by SAW$_1$ (SAW$_2$ or SAW$_3$).}
\label{Fig3}
\end{figure}

The multiplexing action is illustrated by the PL images of Fig.~\ref{Fig3}. Figure~\ref{Fig3}(a) shows the direct transfer of IXs by SAW$_1$ from I/O$_1$ to I/O$_4$, which terminates close to the gate of EXAT$_4$ (kept in the OFF state). The transfer of IXs from SAW$_1$ to either SAW$_2$ or SAW$_3$ is demonstrated in Figs.~\ref{Fig3}(b) and~\ref{Fig3}(c), respectively.  When SAW$_2$ (SAW$_3$) is switched ON, IXs transported by SAW$_1$ to the intersection area are captured by the moving potential dots and transferred to SAW$_2$ (SAW$_3$).  While the dot propagation paths are not spatially resolved in the experiment, their oblique propagation (cf.~Fig.~\ref{Fig2}(c)) is clearly observed within the intersection areas of the SAW beams in Figs.~\ref{Fig3}(b) and~\ref{Fig3}(c).  The few IXs remaining within SAW$_1$ lead to the weak recombination observed in the lower part of this beam. From the PL image in Fig.~\ref{Fig3}(b), we determined that more than 90$\%$ of the IXs are transferred from SAW$_1$ to SAW$_2$ and recombine near EXAT$_3$, thus demonstrating a high beam-to-beam transfer efficiency. These results demonstrate that IXs can be reversibly transfered  from any input port to any output port  by appropriately switching on at most three SAW beams. The overhead time for switching will be established by the bandwidth of the IDTs and by the length of the SAW channels. 

The PL images of Figs.~\ref{Fig3}(a)-\ref{Fig3}(c) have been recorded with the EXATs at the output ports in the OFF state, thus leading to the recombination of the transported IXs close to their gate electrodes. The combined action of long-range IX transport with electrostatic EXAT switching is exemplified in Fig.~\ref{Fig3}(d). 
The image was recorded under the same conditions as Fig.~\ref{Fig3}(c) except that EXAT$_6$ was turned ON by changing its gate bias. Under these conditions, the IXs are transported to the transistor source area. Flow control by the EXAT gate electrodes, which can isolate the sources from the common drain, becomes particularly important for the electric control of IX storage and recombination using the source bias. 

\section{Discussions and Conclusions}
\label{Discussions_and_Conclusions}

The total transport distance between the input and output ports in Fig.~\ref{Fig3}(d) approaches 1000~$\mu$m, thus yielding IX lifetimes exceeding 380~ns. These transport distances, which are well beyond the state of the art in the field, are limited by the length of the transport channel rather than by the acoustic transport efficiency (see Sec.~\ref{Acoustic_exciton_transport}). 
In fact, IX lifetime can be increased by at least another order of magnitude, thus allowing much longer transport lengths with minimal signal loss and distortion, which is essential for the scalability of exciton-based circuitry.

A further interesting feature of the SAW-induced transport is that  the IXs remain trapped   within the 50~$\mu$m-wide  beams (outlined by vertical dotted rectangle in Fig.~\ref{Fig3}(a)). 
The  fabrication of the transport channels does not require lateral electrode structuring or the introduction of lateral interfaces, which can deteriorate the material’s properties. 
Moreover the EXAM concept is  extensive to other systems and can be applied to interconnect an arbitrary number N of remote communication ports. In particular, that the maximum required acoustic transport length (and IX lifetime) does not limit the scalability since it increases only with $\sqrt{N}$. Also, by decreasing the overlap length of the IDT fingers, the width of the SAW beams, and hence of the transport channels, can be further reduced. For a  given length, the minimum channel width will be ultimately determined by SAW diffraction. The latter can be minimized by reducing $\lSAW$ or by shaping the SAW beams using focusing IDTs.
All these features make the acoustic transport a very attractive technique to interconnect remote exciton systems.


In conclusion, the integrated exciton device presented in this work delivers a proof-of-concept for a scalable optoelectronic circuit capable of performing multiple electronic operations using IXs as operation medium. The EXAM is a building block, which can be easily integrated with other devices for on-chip optoelectronic applications. Finally, the long-range IX transport by moving acoustic potentials can provide a versatile framework for the manipulation and coupling of other exciton phases, including IXs or exciton-polaritons condensates.

 {\bf Acknowledgements.} The authors thank M. Ramsteiner for scientific discussions and S. Rauwerdink and W. Seidel for technical support in sample processing. We gratefully acknowledge financial support by DFG project No. SA 598/9.


\end{document}